\documentclass[twocolumn]{aastex63}
\usepackage{newtxtext,newtxmath}
\usepackage[T1]{fontenc}
\usepackage{ae,aecompl}
\usepackage{graphicx}
\usepackage{amsmath}
\usepackage{amssymb}
\usepackage{makecell}

\newcommand{\beq}{\begin{equation}}
\newcommand{\eeq}{\end{equation}}
\newcommand{\beqa}{\begin{equation}\begin{aligned}}
\newcommand{\eeqa}{\end{aligned}\end{equation}}

\newcommand{\Msun}{\, \rm M_\odot}
 
\newcommand{\Mvir}{M_{\rm vir}}

\newcommand{\hiMpc}{\,h^{-1}\rm Mpc}
\newcommand{\hiGpc}{\,h^{-1}\rm Gpc}

\newcommand{\zmin}{z_{\rm min}}

\newcommand{\vpec}{v_{\rm pec}}

\newcommand{\kms}{\,{\rm km\ s^{-1}}}

\newcommand{\vpar}{v_\parallel}
\newcommand{\LCDM}{$\Lambda$CDM}
\newcommand{\scriptM}{\mathcal{M}}

\begin{document}

\title[Peculiar velocities of SNIa]{Specific Effect of Peculiar Velocities on
  Dark-energy Constraints from Type Ia Supernovae}
\author{Dragan Huterer}
\affiliation{Department of Physics, University of Michigan, 450 Church St, Ann
  Arbor, MI 48109-1040, U.S.A.}
\affiliation{Leinweber Center for Theoretical Physics, University of Michigan,
  450 Church St, Ann Arbor, MI 48109-1040}
\affiliation{Max-Planck-Institut f\"ur Astrophysik,
Karl-Schwarzschild-Str.\ 1, 85748 Garching, Germany}

%\pubyear{2020}
\date{\today}% It is always \today, today,

\begin{abstract}
Peculiar velocities of Type Ia supernova (SNIa) host galaxies affect the
dark-energy parameter constraints in a small but very specific way: the
parameters are biased in a single direction in parameter space that is
a-priori knowable for a given SNIa dataset. We demonstrate the latter fact
with a combination of inference from a cosmological N-body simulation with
overwhelming statistics applied to the Pantheon SNIa data set, then confirm 
it by simple quantitative arguments. We quantify small modifications to the
current analyses that would ensure that the effect of cosmological parameters
is essentially guaranteed to be negligible.

\end{abstract}

%%%%%%%%%%%%%%%%%%%%%%%%%%%%%%%%%%%%
\section{Introduction}

Peculiar velocities complicate the Hubble diagram of Type Ia supernovae
(SNIa). A SNIa host galaxy with the peculiar velocity $\mathbf{v}$ will affect
the observed apparent magnitude of the supernova, shifting the observed
redshift to $(1 + z_\text{obs}) = (1 + z)(1 + v_\parallel /c)$
(e.g.\ \cite{Huterer_no_evid}), where $z$ and $z_\text{obs}$ are the true and
observed redshift, and $v_\parallel$ is the peculiar velocity projected along
the line of sight. It is often convenient to recast the shift in SNIa
redshifts to that on the SNIa magnitudes; the latter is (assuming hereafter
that $v_\parallel /c\ll 1$ and $z\ll 1$)
\begin{equation}
  \delta m \simeq \frac{5}{\ln 10}\frac{(1+z)^2}{H(z)d_L(z)}\frac{\vpar}{c},
  \label{eq:dm}
\end{equation}
where $d_L$ is its luminosity distance, $H(z)$ is the Hubble parameter,
and $\vpar$ is the component of peculiar velocity parallel to the line of
sight. For a SNIa at $z\simeq 0.01$ and peculiar velocities of
  order $\vpar\simeq 150$ km/s, this is a shift of $\delta m\simeq (5/\ln
10)(\vpar/cz)\simeq 0.1$ mag, though
  rapidly decreasing with increasing SNIa distance.

The effect of peculiar velocities on cosmological inferences from SNIa has
long been recognized, as it would shift the inferred cosmological parameters,
notably the matter and dark energy densities relative to critical, $\Omega_M$
and $\Omega_\Lambda$, and the dark-energy equation of state parameter $w$
(e.g.\ \cite{HuiGreene06,Davis_2011,CoorayCaldwell06}). To ameliorate the
effect of peculiar velocities on cosmology, early papers advocated adding a
peculiar velocity dispersion of $\simeq (300-400)\kms$ to the magnitude error
\citep{Riess_1998,Perl_1999,Riess_2004,Astier_2006,WoodVasey_2007,Kowalsky_2008,Kessler_2009,Sullivan_2011},
in addition to cutting out the lowest-redshift ($z\lesssim 0.015$) supernovae
from the analysis. A more principled and effective approach is to explicitly
model the covariance of SNIa due to peculiar velocities
\citep{Gorski_1988,Sugiura_1999,Bonvin_2006,HuiGreene06,CoorayCaldwell06},
thus adding such ``signal'' covariance to the ``noise'' contribution from a
combination of statistical and observational systematic errors. The
covariance-matrix approach has been pioneered in the SNIa analysis by \cite{Conley_2011},
and has been adopted, with some variation in the implementation, in most
subsequent analyses
\citep{Davis_2011,Betoule_2014,Scolnic14,Jones_2018,Brout_2019}. Most of the
recent SNIa analyses additionally attempt to remove the effect of peculiar velocities
by using measurements
(e.g.\ \cite{Hudson_Smith_Lucey_2004,Lavaux_Hudson_2011}) from
velocity-reconstructed maps of the nearby large-scale structures.

Previous work has found that, once the triple-pronged measures of removing the
lowest redshift objects, correcting the nearby supernovae for bulk flows, and
modeling the full velocity covariance are enacted, peculiar velocities do not
appreciably affect dark-energy inferences from SNIa (e.g.\ \cite{Davis_2011,Scolnic14}). Nevertheless, this has
occasionally been called into question, often based on suspicions about a
possible large ``bulk flow'' that may lead to larger-than-expected
peculiar-velocity effects \citep{Mohayaee_2020}. While overwhelming evidence
shows that dark-energy constraints of SNIa are not fundamentally changed by
peculiar velocities \citep{Rubin_2016,Rubin_2020} and that SNIa alone indicate
that the velocity field is in agreement with \LCDM\ expectations \citep{Feindt_2013}, it is possible that a
smaller effect of peculiar velocities on dark-energy constraints remains.
%especially given the uncertainties about the structure
%of the local density field.

In this work we calculate, for the first time to our knowledge, the effect of
peculiar velocities on dark-energy constraints evaluated directly from a
numerical ``N-body'' simulation. This approach is robust against any
assumputions about isotropy of either the SNIa distribution or
peculiar-velocity field, and explicitly circumvents any assumptions about the
quality of the velocity-field reconstruction. The only assumption is that the
dark-matter halo velocities trace that of SNIa host galaxy velocities --- that
is, that the velocity bias is equal to unity; this is expected to hold to an
excellent accuracy for our purposes \citep{Wu13bv,Armitage_2018}. To be
maximally conservative, we add the full effect of simulation-inferred peculiar
velocities on the current SNIa data but neglecting any peculiar-velocity
corrections that could be made. Incorporating the latter, as is the practice
in contemporary SNIa analyses, would further ameliorate the effect of peculiar
velocities presented in this work, though to an extent that has not yet been
quantified in detail. We now describe our procedure in detail.

\section{Methodology}

\subsection{Fiducial SNIa data and cosmological models}

We adopt the Pantheon set of 1048 SNIa covering the redshift range
$0.01<z<2.26$ \citep{Pantheon}. We use the full covariance of SNIa magnitude
measurements\footnote{\url{https://github.com/dscolnic/Pantheon}}, which
consists of both noise and signal. In order to test the sensitivity of our
results to the way the covariance has been implemented, we have experimented
with removing the off-diagonals in the published Pantheon covariance, then
adding back corresponding values from our own calculation (specifically,
Eq.~(3.2) in \citet{Huterer_fs8}). We find that the results are robust with
respect to such variations. Of course, ignoring the velocity covariance
altogether would lead to larger effects of peculiar velocities; we emphasize
that all of our results refer to the case when the SNIa covariance has been
fully implemented.

We consider two cosmological models: 1) the curved \LCDM\ model defined with
energy densities of matter and dark energy, $\Omega_M$ and $\Omega_\Lambda$
(as well as the nuisance Hubble-diagram shift parameter $\scriptM$), and 2)
the flat wCDM model defined by $\Omega_M$ (where $\Omega_\Lambda=1-\Omega_M$),
constant equation of state of dark energy $w$, and $\scriptM$. We compute the
fiducial constraints using \texttt{CosmoMC} \citep{cosmomc}. The fiducial
constraints, with errors quoted around the mean, are $\Omega_M=0.319\pm 0.071,
\Omega_\Lambda=0.73\pm 0.11$ (curved \LCDM), and $\Omega_M=0.339\pm 0.064,
w=-1.24\pm 0.24$ (flat wCDM).

\subsection{Peculiar velocities from N-body simulation} \label{sec:dHloc}

To measure and incorporate the effect of peculiar velocities from an N-body
simulation, we repeat a very similar procedure to that in
\cite{Wu_Huterer}. We place an observer at different locations in a large
simulation, identify the closest-matching halos to actual SNIa host galaxy
locations in the Pantheon sample, and directly calculate the effect of peculiar
velocities on cosmological inferences.

Specifically, we use the public release of the Dark Sky
simulations\footnote{\href{http://darksky.slac.stanford.edu}{http://darksky.slac.stanford.edu}}
\citep{Skillman14}, which is run using the adaptive tree code {\tt 2HOT}
\citep{Warren13}. The cosmological parameters correspond to a flat
\LCDM\ model and are consistent with \textit{Planck} and other probes
\cite[e.g.,][]{Planck15Cosmo}: $\Omega_M$ = 0.295; $\Omega_b$ = 0.0468;
$\Omega_\Lambda$ = 0.705; $h$ = 0.688; $\sigma_8$ = 0.835. The dark-matter
halos are identified using the halo finder {\tt Rockstar}
\citep{Behroozi13rs}, and we adopt the largest volume {\tt ds14\_a} with
$10240^3=1.07\times 10^{12}$ particles within $(8 \hiGpc)^3$.
%We use dark matter halos with virial mass $\Mvir > 10^{12.3}\Msun$ (35
%particles).

%DH increase vert spacing in tables
\renewcommand{\arraystretch}{1.2}
%\begin{widetext}
\begin{table*}[htbp]
\centering
\begin{tabular}{l|ccc|ccc}
  \hline
  &  \multicolumn{6}{c}{Cosmological-parameter biases}\\
  & \multicolumn{3}{c}{Curved $(\Omega_M, \Omega_\Lambda)$ model}
  & \multicolumn{3}{c}{Flat $(\Omega_M, w)$ model}\\
  \hline
  
%DH increase vert in this row only
\rule[-.5em]{0pt}{2em}
  $\mathbf{\zmin}$ ($N_{\rm SN}$)
& Median $\Delta\chi^2_{\rm 2d}$
& $p\left (\Delta\chi^2_{\rm 2d}> 2.3\right )$
& Relative  FoM
& Median $\Delta\chi^2_{\rm 2d}$
& $p(\Delta\chi^2_{\rm 2d}> 2.3)$
& Relative  FoM
\\
\hline
0.01 (1048)
& $0.38 $ & $9.7\%$ & $1.0$ 
& $0.48 $ & $14.2\%$& $1.0$ \\
\hline
0.02 (1002)
& $0.30 $ & $5.8\%$ & $0.99$ 
& $0.37  $ & $8.7\%$ & $0.95$ \\
\hline
0.03 (953)
& $0.14 $ & $0.55\%$ & $0.94$ 
& $0.17 $ & $1.25\%$ & $0.89$ \\
\hline
\end{tabular}
\caption{Summary of the statistics of shifts in the (projected)
  two-dimensional dark-energy-parameter plane of interest. We show the median
  shift in $\chi^2$, the percentage of realizations that have $\Delta\chi^2$
  greater than 2.3 (the Gaussian 68\% value), and the relative figure-of-merit
  (inverse area of the 2D contour) in the given parameter plane.}
\label{tab:dchisq}
\end{table*}
\renewcommand{\arraystretch}{1.0}

We then divide this $(8\hiGpc)^3$ volume into 512 subvolumes of $(1\hiGpc)^3$,
and consider halos with virial mass $\Mvir \in [10^{12.3}, 10^{12.4}]\Msun$
(which is roughly the Milky Way mass). In each subvolume, we first identify
the halo that is closest to the center; this will be the location of that
subvolume's observer.  Relative to this observer location, we then find the
closest halo to each Pantheon SNIa location in space (a halo in our mass range
can be typically find within $\sim15\hiMpc$ of a given 3D location).

%% that is closest to the center of each subvolume.  This choice
%% simulates 512 separate observers located on Milky-Way-mass halos, each with a
%% separate subvolume of the large-scale structure. In each subvolume, we find the halo
%% closest to the center . We place the observer on this halo which,
%% given our choice for halo masses, is roughly the Milky Way mass.

While the relative positions of SNIa in redshift and angle are fixed, the orientation
of their coordinate system relative to that of the simulation frame is
arbitrary and, given the highly inhomogeneous distribution of SN in the
volume, may likely lead to additional variance. To account for this, we
explore all possible orientations of the SNIa frame relative to the (fixed)
subvolume frame. To vary over the orientations, we employ 3240 Euler angles;
see the Appendix of \citet{Wu_Huterer} for details. We therefore have a sample of
$512\times 3240$, or around 1.65 million, realizations of peculiar-velocity
field centered at an observer. In each realization, the radial velocity of
SNIa is given simply by
\begin{equation}
v_{\parallel,i}\equiv \textbf{v}_i\cdot
\frac{(\textbf{r}_i-\textbf{r}_{\rm obs})}{|\textbf{r}_i-\textbf{r}_{\rm obs}|} \ ,
\end{equation}
where $\textbf{r}_i$ and $\textbf{v}_i$ are the location and velocity of the
closest halo to the $i^{\rm th}$ SNIa, and $\textbf{r}_{\rm obs}$
is the location of the observer. We add this peculiar velocity to Pantheon SNIa
magnitudes at\footnote{We do this to speed up calculations, and have checked that the
  results are completely converged for this redshift range.} $z<0.1$ by employing Eq.~(\ref{eq:dm}). We then
 calculate the cosmological biases as described in the next subsection.

Because we are looking at the relative \textit{change} in the effective
redshift of observed SNIa, we assume that the measured $\vpec$ of their host
halos has been unaccounted for and is to be added to the (CMB rest-frame)
redshift measured in Pantheon objects.  As mentioned in the introduction, this
assumption is clearly conservative, as it assumes that the misestimate of the
peculiar velocity is large, being equal to its full value of $\vpec$ for each
SNIa (the expected error is presumably some fraction of the measured
$\vpec$). 

%%%%%%%%%%%%%%%%%%%%%%%%%%%%%%%%%%%%
\subsection{Cosmological-parameter bias calculation}\label{sec:discussions}

We finally need to calculate the biases in the cosmological parameters given
peculiar-velocity shifts in SNIa magnitudes.  To do that, adopt the Fisher
matrix bias formula \citep{Knox_1999,Huterer_2002} and calculate the
linearized shift in the cosmological parameters $\{p_i\}$ given the change in
the redshifts due to peculiar velocities:
\begin{equation}
  \delta p_i\approx (F^{-1})_{ij} \sum_{a, b} \delta(m)_a\,
  \mathbf{C}[m(z_a), m(z_b)]^{-1}\,
{\partial \bar m(z_b) \over \partial p_j}, 
\label{eq:Fisherbias}
\end{equation}
where $\delta m$ is the magnitude shift due to peculiar velocity given in
Eq.~(\ref{eq:dm}), $\mathbf{C}$ is the full SNIa data covariance, and
$\partial \bar m(z_b)/\partial p_j$ is the sensitivity of the
theoretically computed magnitude to shifts in cosmological
parameters. Finally, $F$ is the Fisher matrix (approximation of the inverse parameter
covariance matrix) for the distribution of SNIa in Pantheon and the three
cosmological parameters that are, recall, $(\Omega_M, \Omega_\Lambda,
\scriptM )$ and $(\Omega_M, w, \scriptM )$, respectively, for the two models
that we study. The Fisher bias formula is expected to be accurate in the limit
of small shifts, which is what we have at hand. We explicitly tested its
accuracy by recomputing the full cosmological constraints for the case when
SNIa magnitudes are shifted by peculiar velocities in a few numerical
realizations, and verified that the final shifts in the best-fit values in the
parameters $\{p_i\}$ are accurately approximated by
Eq.~(\ref{eq:Fisherbias}). Finally, for each peculiar-velocity realization, we
evaluate $\Delta\chi^2_{\rm 2d}=\mathbf{(\delta p)}^T \mathbf{F}_{\rm
  2x2}^{-1}\mathbf{(\delta p)}$, where $\mathbf{\delta p}$ is the length-two
vector of biases in the two parameters, and $\mathbf{F}_{\rm 2x2}$ is the
Fisher matrix projected to the relevant two-dimensional parameter space.

\section{Results}

We perform the analysis described above for three cases of the minimum
redshift of SNIa: $\zmin=0.01$ (which contains all 1048 Pantheon SNIa),
$\zmin=0.02$ (1002 SNIa), and $\zmin=0.03$ (953 SNIa). As $\zmin$ increases,
the statistical results slightly weaken, but the peculiar-velocity-induced
biases dramatically decrease. We wish to study the interplay between the two
effects, with the desired goal to keep as many of the low-z objects as
possible \citep{Linder_lowz_imp}.

The results are shown in Table \ref{tab:dchisq}. For each of the three $\zmin$
choices, and for both $(\Omega_M, \Omega_\Lambda )$ and $(\Omega_M, w)$
parameter space, we show the median $\Delta\chi^2_{\rm 2d}$, the percentage of
realizations that have $\Delta\chi^2_{\rm 2d}$ greater than 2.3 (the Gaussian 68\%
value), and the relative figure-of-merit (inverse area of the relevant 2D
contour). We observe that the biases (quantified by $\Delta\chi^2_{\rm 2d}$) start
out nonnegligible, but dramatically decrease with $\zmin$, while the FoM
decreases very slightly from $\zmin=0.01$ to $0.02$, and somewhat more but
still modestly from 0.02 to 0.03.

\begin{figure*}
    \centering
        \includegraphics[width=0.48\linewidth]{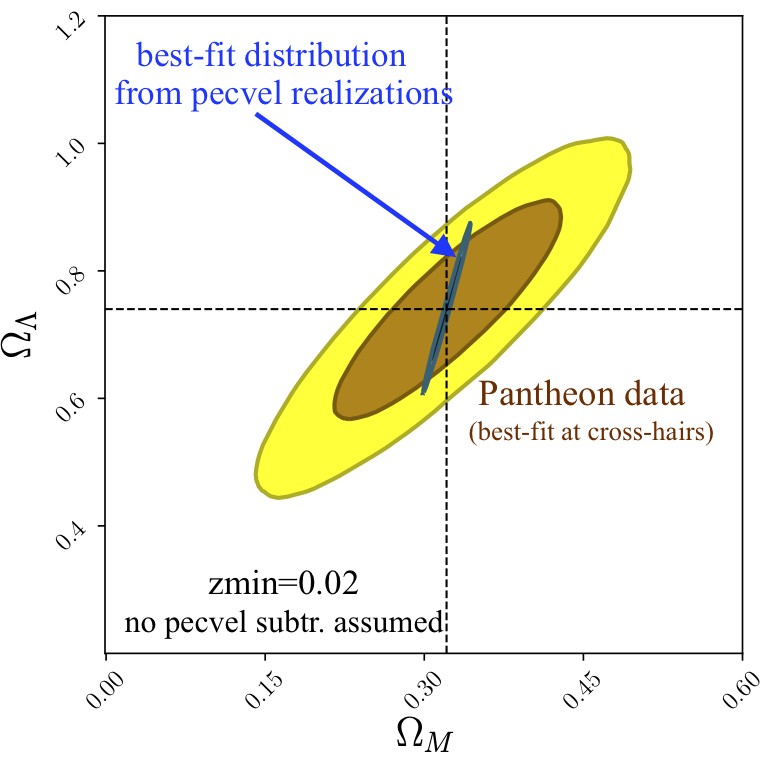}
        \includegraphics[width=0.49\linewidth]{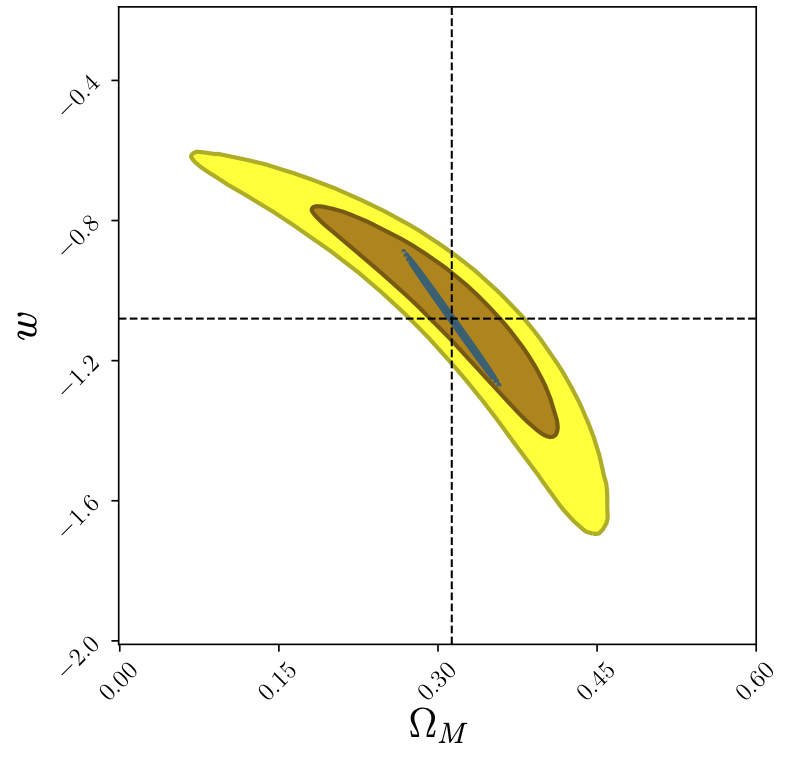}
    \caption{Yellow contours show the fiducial 68.3\% and 95.4\%
      C.L.\ constraints on $(\Omega_M, \Omega_\Lambda)$ (left panel) and
      $(\Omega_M, w)$ (right panel) with the fiducial Pantheon SNIa
      dataset. The blue contours show the 68.3\% and 95.4\% intervals of the
      shifts of the best-fit model (whose central value --- the mean from our
      chains --- is shown by the cross-hairs), evaluated from 1.65 million
      realizations of the peculiar-velocity field. In this plot we use SNIa with
      $z>0.02$. See the text for other details.}
    \label{fig:plot}
\end{figure*}

The results for both models and for $\zmin=0.02$ are pictorially shown in
Fig.~\ref{fig:plot}.  The fiducial 68.3\% and 95.4\% constraints from
Pantheon are given with the two larger set of contours, with the best-fit (mean value
from the chains) given in the crosshairs. The smaller set of contours in each panel describes
the distribution of shifts of the best-fit value in cosmological parameters due to
peculiar velocities. Recall that there are 1.65 million such realizations;
they are distributed with mean very close to zero (so that the \textit{mean} peculiar-velocity realization does not affect the fiducial Pantheon
analysis), and spread described by two contours that describe the 68.3\% and
95.4\% mass of the parameter shifts.

Moreover, Fig.~\ref{fig:plot} dramatically illustrates that the contours have
a very specific direction in either 2D plane. The direction is given by
\begin{equation}
    \begin{aligned} 
  \delta\Omega_\Lambda &= +5.7\,\delta\Omega_M \qquad \mbox{(curved $\Omega_M-\Omega_\Lambda$)}\\[0.0cm]
  \delta w &= -4.3\, \delta\Omega_M \qquad \mbox{(flat\qquad $\Omega_M-w$)},
    \end{aligned}
    \label{eq:monopole_biases}
\end{equation}
where the coefficients vary slightly as a function of $\zmin$; the above
values are for $\zmin=0.02$.  Inspection of Fig.~\ref{fig:plot} shows that the
one-dimensional approximation to the general biased in the respective 2D spaces
is entirely appropriate.

The fact that the cosmological-parameter biases always lie in (fixed) 1D
directions hints at the fact that the SNIa Hubble diagram is mainly sensitive
to the overall monopole of the peculiar-velocity field --- the overall
sky-averaged gradient centered at the observer.  We explicitly checked this by
calculating the biases in cosmological parameters by artificially shifting the
SNIa Hubble diagram at low z ($z\lesssim 0.03$) by an arbitrary but
redshift-independent amount $\delta m$ and repeating the parameter-bias
calculation using Eq.~(\ref{eq:Fisherbias}). The resulting bias agrees well
with Eq.~(\ref{eq:monopole_biases}) and blue contours in Fig.~\ref{fig:plot},
confirming that the bias is driven by the overall divergence of the density
field evaluated at the observer's location (the ``monopole'').

The remaining task is to ensure that the peculiar-velocity effects are
negligible. To that end, we employ the simplest strategy of simply adjusting
the minimal redshift in order to suppress the bias while ensuring a healthy
population of low-z SNIa that are required to have excellent cosmological
constraints. Results in Table \ref{tab:dchisq}
indicate that $\zmin=0.02$ is sufficient to protect against biases. Perhaps a
more realistic estimate of the biases in current SNIa analyses is obtained
assuming that \textit{half} of the peculiar velocities are removed by the
velocity-field-correction techniques. In that case, and for
 $\zmin=0.02$, the fraction of cases when $\chi^2_{\rm 2d}>2.3$ reduces to
mere 0.07\% and 0.02\% for the two respective cosmological models.

%%%%%%%%%%%%%%%%%%%%%%%%%%%%%%%%%%%%
\section{Discussion and Conclusions}\label{sec:conclusions}

We have evaluated the effect of (uncorrected-for) peculiar velocities on the
cosmological constraints from the Pantheon SNIa dataset. Our inference is
based on a large-volume N-body simulation with more than 1.5 million
realizations of the local peculiar-velocity field.  Given these overwhelming
statistics, we are confident that our procedure accurately reflects the
ensemble of peculiar-velocity fields allowed within \LCDM, and that any
purported cases of a ``local void'' are captured within the statistical
distribution. The two assumptions we have made are: 1) that the
cosmological model is the current \LCDM\ (which is at least approximately true given
excellent constraints from non-SNIa data), and 2) that the velocity bias ---
velocities of galaxies relative to those of dark-matter halos --- is close to
unity, which is also  supported by independent work.

We find that the peculiar velocities bias the cosmological parameters in very
specific 1D directions; see Fig.~\ref{fig:plot}. This, in turn, indicates that
the dominant effect of peculiar velocities is their overall monopole relative
to the observer.  We back up this conjecture with a simple numerical experiment.

No subtraction of peculiar velocities was assumed in our procedure. Therefore,
the results shown in Table \ref{tab:dchisq} and Fig.~\ref{fig:plot} represent a conservative estimate
of the effects of peculiar velocities. They indicate that, even under these
conservative assumptions, removing $z\lesssim 0.02$ SNIa, along with the
``good-health habit'' of including the full velocity covariance, leads to
negligible biases in the cosmological parameters.

\section*{Acknowledgments}
We thank Heidi Wu for earlier collaboration  on \cite{Wu_Huterer}, and Alex
Kim, Eric Linder, and Fabian Schmidt for useful feedback on the manuscript. DH
is supported by DOE, NASA, NSF, and Alexander von Humboldt Foundation. We analyzed the MCMC chains and plotted their
results using \texttt{ChainConsumer} \citep{ChainConsumer}.

\bibliography{refs}

\end{document}